\documentclass[prd,twocolumn,showpacs,floatfix,amsmath,nofootinbib,amssymb,floatfix]{revtex4}
\usepackage{graphicx,color,dcolumn,booktabs,bm}
\usepackage{longtable,lscape}
\usepackage{txfonts}
\usepackage{overpic}
\usepackage{amssymb}
\usepackage{indentfirst}
\usepackage{feynmf}   
\usepackage{slashed}  
\usepackage{cases}
\usepackage{color}
\usepackage{multirow}
\usepackage{epstopdf}
\usepackage{longtable}
\usepackage{graphicx,color,dcolumn,booktabs,bm}
\usepackage[colorlinks,
            citecolor=blue,
            anchorcolor=red,
            menucolor=red,
            linkcolor=red,
            filecolor=red,
            runcolor=red,
            urlcolor=blue,
            frenchlinks=red]{hyperref}

\graphicspath{{Figures/}} %

\allowdisplaybreaks

\begin{document}

\title{Explanation of Y(4630) as hadronic resonant state}

\author{Xiao-Hui Mei$^{1}$}
\author{Zhuo Yu$^{2}$}
\author{Mao Song$^{1}$}
\email{songmao@mail.ustc.edu.cn}
\author{Jian-You Guo$^{1}$}
\author{Gang Li$^{1}$}
\author{Xuan Luo$^{1}$}

\affiliation{ $^1$School of Physics and Optoelectronic engineering, Anhui University, Hefei 230601, China
\\
$^2$School of Physics, Southeast University, Nanjing 210094, China}

\begin{abstract}
After Y(4630) is discovered, theorists have given various explanations. We find that if Y(4630) is interpreted as the D-wave resonant state of $\Lambda_c \bar {\Lambda}_c$ system, the particle mass, decay width and all quantum numbers are consistent with experimental observations. We use the Bonn approximation to get the interaction potential of one boson exchange model, then extend the complex scaling method (CSM) to calculate the bound and resonant states. The results indicate that the $\Lambda_c \bar{\Lambda}_c$ system can form not only the bound state of S wave, but also the resonant state of the high angular momentum, and the $^3D_1$ wave resonant state can explain the structure of Y(4630) very well.
\end{abstract}


\maketitle

\section{introduction}\label{sec1}

Deuteron is a proton and neutron molecular state, which is well explained by one boson exchange model \cite{Tornqvist:1993vu,Tornqvist:1993ng}.
Along this line, we may wonder whether the heavy baryon pair can also form a deuteron-like bound state by exchanging virtual light mesons.
Intuitionally, the larger mass of the heavy baryons can reduce the kinetic of the systems and easier to form bound states.
Therefore, it is interesting to study whether one boson exchange interactions are strong enough to bind the two heavy baryons (dibaryon) or a heavy baryon and an anti-baryon (baryonium).

\par
$\Lambda_c$ is the lightest charmed baryon, which contains a charm quark and two light quarks, and the composition is similar to proton.
Having more knowledge about $\Lambda_c$ is helpful for studying the properties of other charmed baryons. Until now, our understanding about the $\Lambda_c$ behavior is very limited. The Belle Collaboration firstly reported a charmonium-like state
$Y(4630)$ in $\Lambda_c \bar{\Lambda}_c$ invariant mass spectrum from the $e^+e^- \to \gamma_{ISR} \Lambda_c \bar{\Lambda}_c$ process, where $\gamma_{ISR}$ is the emitted photon from the initial leptons, the related parameters are mass $M = 4634^{+8+5}_{-7-8}$ MeV, width $\Gamma = 92^{+40+10}_{-24-21}$ MeV, quantum number $J^{PC}=1^{--}$ \cite{Pakhlova:2008vn}.

After the observation of $Y(4630)$, various theoretial interpretations were proposed, such as conventional charmonium state~\cite{Badalian:2008dv,Segovia:2008ta}, tetraquark state~\cite{Maiani:2014aja,Cotugno:2009ys,Brodsky:2014xia}, $\Lambda_c \bar \Lambda_c$ baryonium~\cite{Lee:2011rka,Chen:2011cta} and threshold effect~\cite{vanBeveren:2008rt}. Simonov proposed a mechanism to study
baryon-antibaryon production, which can explain why the $Y(4630)$ enhancement structure appears in the electroproduction of $\Lambda_c \bar\Lambda_c$
\cite{Simonov:2011jc}. Recently, a series of investigations on the strong decay behaviors were proposed, which intended to reveal the inner structure of  $Y(4630)$~\cite{Liu:2016sip,Liu:2016nbm,Guo:2016iej}.

Resonance is one of most striking phenomenon in the whole range of scattering experiments, which appear widely in atoms, molecules, nuclei and chemical reactions.  Based on conventional scattering theory, $R$-matrix method \cite{Wigner:1947zz, Hale:1987zz}, $K$-matrix method \cite{Humblet:1991zz}, scattering phase shift method, continuous spectrum theory, $J$-matrix method \cite{Taylor} have been developed. For the convenience of calculation, several bound-state-like methods are developed, such as real stabilization method (RSM)\cite{Hazi}, analytic continuation method of coupling constant (ACCC) \cite{Kukulin} and complex scaling method (CSM) \cite{csm1,csm2}, $etc$. The complex scaling method can describe the bound state, resonant state and continuum in a consistent way, which is widely used to exploring the resonance in atomic, molecular and nuclear physics. The CSM has been extended from nonrelativistic to the relativistic framework \cite{csm3,csm4,csm5,csm6,csm7}and from spherical nuclei to deformed nuclei \cite{csm9}, which has been applied in halo nuclei. In Ref.\cite{Yu:2021lmb}, the authors firstly extend CSM from atomic, molecular and nuclear physics to hadron physics for explaining hadron molecular state.
Recently, the CSM has been used more and more in hadronic physics \cite{Wang:2022yes,Cheng:2022qcm,Cheng:2022vgy}.

Among various explanations, hadronic molecules gain more attention, since Y(4630) is close to the thresholds of two hadrons.
The bound state of S wave is easier to form, but the higher excited states also have a certain probability to form.
For example, the heavy quarkonium not only found the ground state $J/\psi$, but also the excited states $h_c(1P)$, $\chi_{c2}(2P)$ and $\Upsilon(1^3D_2)$ were observed in experiment. The higher excited states of hadrons provide a unique way to study the structure of hadron states.
In the framework of one-boson-exchange model, if the hadrons can bind to hadron molecular states, whether can form resonant states with high angular momentum. In Ref.\cite{Yu:2021lmb}, the authors have calculated the resonant states for $DD(\bar{D})$, $\Lambda_cD(\bar{D})$ and $\Lambda_c\Lambda_c(\bar{\Lambda}_c)$ systems in heavy quark effective theory. The bound state of $\Lambda_c\bar{\Lambda}_c$ system has been investigated in several previous works \cite{Chen:2017jjn,Meguro:2011nr,Li:2012bt,Lee:2011rka}. In this paper, we will consider the spin-orbit coupling effect and further investigate whether $\Lambda_c \bar{\Lambda}_c$ system can form a resonant state consistent with the quantum number, mass and width of Y(4630).

This paper is organized as follows. After the introduction, we present the theoretical framework and calculation method in Section \ref{sec2}. The numerical results and discussion are given in Section \ref{sec3}. A short summary is given in Section \ref{sec4}.

\section{Theoretical framework}\label{sec2}
In our work, we calculate the effective interaction potential for the $\Lambda_c\bar{\Lambda}_c$ system in the Bonn meson-exchange model.
Due to the spin and isospin conservation in hadron systems, the contributions of $\pi$, $\eta$, $\rho$ meson exchanges are forbidden or suppressed heavily.
The interactions of $\Lambda_c\bar{\Lambda}_c$ system are mainly mediated by $\sigma$ and $\omega$ mesons, and
the effective Lagrangian densities for one-$\sigma$-exchange and one-$\omega$-exchange are expressed,

\begin{eqnarray}
\mathcal{L} = g_{\sigma \Lambda_c \Lambda_c}\bar\psi\sigma \psi -g_{\omega \Lambda_c \Lambda_c}\bar\psi
\gamma_{\mu} \omega^{\mu}\psi
+\frac{f_{\omega \Lambda_c \Lambda_c}}{2m_{\Lambda_c}}\bar\psi\sigma_{\mu\nu}\psi
\partial^{\mu}\omega^{\nu},\label{one-exchange}
\end{eqnarray}

Here, $\psi$ is the Dirac-spinor for the spin-$\frac{1}{2}$ particle of $\Lambda_c$. In the tensor coupling term, the constant $f_{\omega \Lambda_c \Lambda_c}=-g_{\omega \Lambda_c \Lambda_c}$ in Ref.\cite{Lee:2011rka}, so the tensor coupling coefficient is $2m_{\Lambda_c}$ lower than vector coupling coefficient. Moreover the tensor term is proportional to the relative momentum $q^\nu$, which is small, thus the contribution of the tensor term can be ignored.

Although there are no definite values for the coupling strengths $g_{\omega/\sigma \Lambda_c \Lambda_c}$ in experiment, they can estimated by using the quark model. Since the exchange of $\sigma$ and $\omega$ mesons occurs mainly between the light quarks in heavy hadrons, the interactions of light quarks $(q=u,d)$ and $\sigma/\omega$ can be written as
\begin{eqnarray}
\mathcal{L}_{qq\sigma/\omega} &=& -g_{\sigma}^{q}\bar{\psi}_q\sigma\psi_q-g_{\omega}^{q}\bar{\psi}_q\gamma^{\mu}\omega_{\mu}\psi_q.\label{quark}
\end{eqnarray}
Compared with the vertices of $\bar{\Lambda}_c\Lambda_c\sigma/\omega$ and $\bar{q}q\sigma/\omega$ in Eqs. (\ref{one-exchange})$-$(\ref{quark}), the coupling constants can be related by,
\begin{eqnarray}\label{cc}
g_{\sigma \Lambda_c \Lambda_c} = 2 g_{\sigma}^{q}, \quad\quad
g_{\omega \Lambda_c \Lambda_c} = 2 g_{\omega}^{q}.
\end{eqnarray}
In a $\sigma$ model \cite{Riska:1999fn}, the value of $g_{\sigma}^{q}$ is taken as $g_{\sigma}^{q}=3.65$. For the $\omega$ coupling $g_{\omega}^{q}$, in the Nijmegen model, $g_{\omega}^{q}=3.45$, whereas it is equal to 5.28 in the Bonn model \cite{Rijken:1998yy}. In Ref. \cite{Riska:2000gd}, $g_{\omega}^{q}$ was roughly assumed to be 3.00.

\begin{figure}[ht]
  \begin{center}
    \rotatebox{0}{\includegraphics*[width=0.4\textwidth]{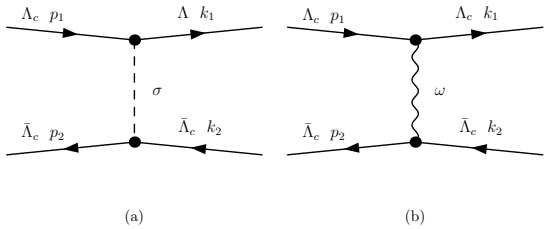}}
    \caption{The Feynman diagrams at the tree level.\label{feynman}}
  \end{center}
\end{figure}

Based on the Lagrangins in Eqs~(\ref{one-exchange}), we can obtain the scattering Feynman amplitudes for $\Lambda_c\bar{\Lambda}_c \to \Lambda_c\bar{\Lambda}_c$ in Fig.~\ref{feynman}. The annihilation effect from S-channel has not been take into account in the calculation. In the center-of-mass frame, the initial four-momenta are $p_1(E_1,\vec{p})$ and $p_2(E_2,-\vec{p})$, the final four-momenta are $k_1(E_1,\vec{p'})$ and $k_2(E_2,-\vec{p'})$, as shown in Fig.\ref{feynman}. Thus the four-momenta of propagator is
\begin{equation}
q=k_1-p_1=p_2-k_2=(0,\vec{p'}-\vec{p})=(0,\vec{q})
\end{equation}
For convenience of calculations, we make the substitution for the following four-momenta,
\begin{equation}
\vec{q}=\vec{p'}-\vec{p}, ~~~\vec{k}=\frac{1}{2}(\vec{p}+\vec{p'}).
\end{equation}

In the nonrealistic approximation, we keep the terms up to order of $\frac{1}{m_{\Lambda_c}^2}$. The scattering
amplitudes are

\begin{eqnarray}
iM_{\sigma}=&&-g_{\sigma \Lambda_c \Lambda_c}^2\bar{u}(k_1)u(p_1)
\frac{i}{q^2-m_{\sigma}^2}
\bar{\upsilon}(p_2)\upsilon(k_2) \nonumber\\
=&&i\frac{g_{\sigma \Lambda_c \Lambda_c}^2}{\vec{q}^{2}+{m_\sigma}^2}
\left[1 - \frac{\vec{k}^{2}}{2 m_{\Lambda_c}^2} + \frac{\vec{q}^{2}}{8
m_{\Lambda_c}^2} + i\frac{\vec{S}\cdot(\vec{k}\times
\vec{q})}{2m_{\Lambda_c}^2}\right],\label{sigma-amplitude}
\end{eqnarray}
and
\begin{eqnarray}
iM_\omega=&&-g_{\omega \Lambda_c \Lambda_c}^2\bar{u}(k_1)\gamma^{\mu}u(p_1)
i\frac{-g_{\mu\nu}+\frac{q_{\mu}q_{\nu}}{m_{\omega}^2}}{q^2-m_{\omega}^2}\bar{\upsilon}(p_2)\gamma^{\nu} \upsilon(k_2)\nonumber\\
=&&i\frac{g_{\omega \Lambda_c \Lambda_c}^2}{\vec{q}^2+m_{\omega}^2}
\Bigg[1-\frac{\vec{q}^{2}}{8m_{\Lambda_c}^2} + \frac{3\vec{k}^{2}}{2
m_{\Lambda_c}^2}+
i\frac{3\vec{S}\cdot\left(\vec{k}\times\vec{q}\right)}{2m_{\Lambda_c}^2}\nonumber\\
&&-\frac{(\vec{\sigma_1}\cdot\vec{\sigma_2})\cdot\vec{q}^2}{4m_{\Lambda_c}^2}
+\frac{(\vec{\sigma}_1\cdot\vec{q})(\vec{\sigma}_2\cdot\vec{q})}{4m_{\Lambda_c}^2}\Bigg],
\label{omega-amplitude}
\end{eqnarray}
where $\vec{S}=\frac{1}{2}(\vec{\sigma}_1+\vec{\sigma}_2)$ is the total spin of $\Lambda_c\bar{\Lambda}_c$ system.

In the Breit approximation, the relation between the effective potential in momentum space $\mathcal{V}_{fi}$ and the scattering amplitude $\mathcal{M}_{fi}$ in the momentum space are expressed,
\begin{eqnarray}
\mathcal{V}_{fi}(\bm{q}) &=& -\frac{\mathcal{M}_{fi}(\Lambda_c\bar{\Lambda}_c \to \Lambda_c\bar{\Lambda}_c)}{\sqrt{\prod_i2m_i\prod_f2m_f}}.
\end{eqnarray}
Here, $m_i$ and $m_f$ are the masses of the initial ($\Lambda_c$, $\bar{\Lambda}_c$) and final particles ($\Lambda_c$, $\bar{\Lambda}_c$), respectively.

In the above paper, hadron is directly treated as point particles without considering the internal structure of the hadron. In order to regularize the off shell effect of the exchanged meson, it is necessary to introduce a monopole form factor $\mathcal{F}(q^2)$ at every vertex, which has a form as:

\begin{eqnarray}
\mathcal{F}(q^2) &=& \frac{\Lambda^2-m^2}{\Lambda^2-q^2},
\end{eqnarray}
here, $\Lambda$ is the cutoff parameter, $m$ and $q$ correspond to the mass and momentum of the exchanged meson, respectively.
In Refs.\cite{Tornqvist:1993vu,Tornqvist:1993ng}, $\Lambda$ is related to the root-mean-square radius of the source hadron which propagate the interaction through the intermediate boson ($\sigma$ or $\omega$). According to the previous experience of the deuteron, the cutoff $\Lambda$ is taken around 1.0 GeV.

After adding the monopole form factor $\mathcal{F}(q^2)$,
the effective potential in the coordinate space $\mathcal{V}(r)$ is obtained by performing the Fourier transformation as
\begin{eqnarray}\label{fourier}
\mathcal{V}(\bm{r}) &=& \int\frac{d^3\bm{q}}{(2\pi)^3}e^{i\bm{q}\cdot\bm{r}}\mathcal{V}(\bm{q})\mathcal{F}^2(q^2).
\end{eqnarray}

The detailed Fourier transformations for different types of effective potentials are expressed as \cite{Chen:2019uvv,Zhao:2014gqa}
\begin{eqnarray}
&\mathcal{F}&\left\{\frac{1}{\vec{q}^2+m^2}\left(\frac{\Lambda^2-m^2}
{\Lambda^2+\vec{q}^2}\right)^2\right\} = Y\left(\Lambda,m,r\right),\nonumber\\
&\mathcal{F}&\left\{\frac{\vec{q}^2}{\vec{q}^2+m^2}\left(\frac{\Lambda^2-m^2}
{\Lambda^2+\vec{q}^2}\right)^2\right\}=-\nabla^2Y\left(\Lambda,m,r\right),\nonumber\\
&\mathcal{F}&\left\{\frac{\vec{k}^2}{\vec{q^2}+m^2}\left(\frac{\Lambda^2-m^2}
{\Lambda^2+\vec{q}^2}\right)^2\right\}\nonumber=\frac{1}{4}\nabla^2Y\left(\Lambda,m,r\right)
\nonumber\\&&-\frac{1}{2}\left\{\nabla^2,Y\left(\Lambda,m,r\right)\right\},\nonumber\\
&\mathcal{F}&\left\{\frac{\left(\vec{S}\cdot(\vec{q}\times\vec{k})\right)}
{\vec{q}^2+m^2}\left(\frac{\Lambda^2-m^2}{\Lambda^2+\vec{q}^2}\right)^2\right\}
=-i\vec{S}\cdot\vec{L}\frac{1}{r}\frac{\partial}{\partial r}Y\left(\Lambda,m,r\right),\nonumber\\
&\mathcal{F}&\left\{\frac{\left(\vec{\sigma}_1\cdot\vec{q}\right)
\left(\vec{\sigma}_2\cdot\vec{q}\right)}{\vec{q}^2+m^2}
\left(\frac{\Lambda^2-m^2}{\Lambda^2+\vec{q}^2}\right)^2\right\}\nonumber\\
&=&-\frac{1}{3}\left(\vec{\sigma}_1\cdot\vec{\sigma}_2\right)
\nabla^2Y\left(\Lambda,m,r\right)-\frac{1}{3}S\left(\vec{\hat{r}},\vec{\sigma}_1,
\vec{\sigma}_2\right)T\left(\Lambda,m,r\right),\nonumber\\
\end{eqnarray}

As shown in Ref.\cite{Lee:2011rka}, due to the cancellations of the coupling constants $f_{\omega \Lambda_c \Lambda_c}$ and $g_{\omega \Lambda_c \Lambda_c}$ in the tensor terms, and there is no mixing of S and D states. Therefore, $\Lambda_c\bar{\Lambda}_c$ systems do not need to consider S-D coupling.
The $\vec{k}^2$ term is named as the recoil correction term, and the function $Y\left(\Lambda,m,r\right)$ is defined as
\begin{eqnarray}
Y\left(\Lambda,m,r\right)&=&\frac{1}{4\pi r}\left(e^{-mr}-e^{-\Lambda r}\right)-\frac{\Lambda^2-m^2}{8\pi\Lambda}e^{-\Lambda r}.
\end{eqnarray}

The one-$\sigma$-exchange and one-$\omega$-exchange interactions are corresponding to intermediate- and short-range forces,
therefore they are suppressed when the radius $r$ reaches 1.0 fm or larger. For the $^1S_0$ state, both the $\omega$-exchange and $\sigma$-exchange
provide attractive force. For the $^3S_1$ state, the vector meson $\omega$ provides repulsive force in the short range
but attractive force in the medium range, while the scalar meson $\sigma$ always provides attractive force.

After the interaction potential $\mathcal{V}(r)$ in coordinate space is obtained, the eigenvalue and eigenfunction of the bound state for $\Lambda_c\Lambda_c$ system can be obtained by solving the non-relativistic Schr\"{o}dinger equation. In this paper, we extend the CSM to solve the Schr\"{o}dinger equation in the complex energy plane. The Aguilar-Balslev-Combes(ABC) theorem \cite{abc} proved that under the complex scaling transformation, the energy spectrum has three parts: (i) the bound states are discrete set of real points on the negative energy axis and remain unchanged under complex scale transformation; (ii) the resonant states correspond to the discrete set of points in the lower half of complex energy plane, which does not change with the coordinate transformation; and (iii) The continuous spectrum is rotated at $2\theta$ around the origin of the coordinates. In the CSM, bound states, resonant states and continuous spectra can be described uniformly.  As the $\theta$ increases, the resonant states are exposed in the fourth quadrant of the complex energy plane and do not change with the rotation of the continuous spectrum. We solve the complex scaled Schr\"{o}dinger equation by basis expansion method, where the radial function use spherical harmonic oscillator basis. The detailed calculation scheme can refer to our previous work \cite{Yu:2021lmb}.

\section{Numerical results}\label{sec3}

In this section, we discuss and analyze the effects of the one-$\sigma$-exchange and one-$\omega$-exchange interactions for $\Lambda_c \bar{\Lambda}_c$ system. The total boson exchange potentials are used to calculate the numerical solution of the Schr\"{o}dinger equation in the CSM. The related parameters are given in Table \ref{parameter}.
\renewcommand\tabcolsep{0.16cm}
\renewcommand{\arraystretch}{1.8}
\begin{table}[!htbp]
  \caption{The related parameters are used in this work \cite{Olive:2016xmw}.}\label{parameter}
  \begin{tabular}{ccc|ccc}\toprule[2pt]
  Hadron     &$I(J^P)$     &Mass (MeV)    &Hadron     &$I(J^P)$     &Mass (MeV) \\\hline
  $\sigma$   &$0(0^+)$              &600          &$\omega$        &$0(1^-)$               &782.65\\
  $\Lambda_c$     &$0(\frac{1}{2}^+)$     &2286.46 &      &   &      \\
  \bottomrule[2pt]
  \end{tabular}
\end{table}
The corresponding eigenvalues can be obtained by diagonalizing the Hamiltonian, then we can get the information about the resonant state.

\begin{figure}
\centering
\includegraphics[width=3.10in, keepaspectratio]{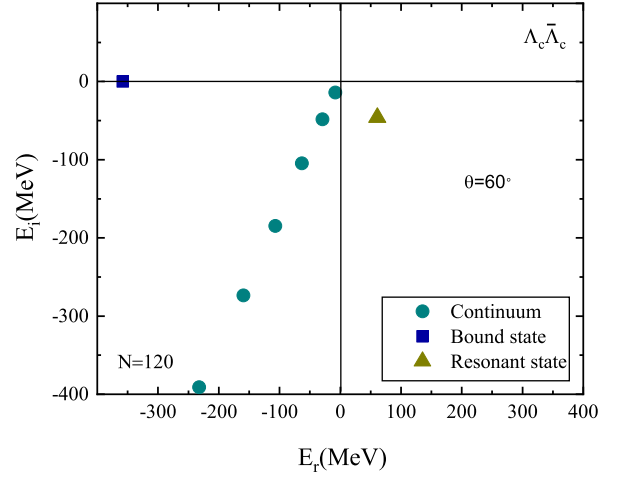}\\
\caption{(Color online) The resonant state is presented with $\theta = 60^{\circ}$. Here, the cutoff parameter $\Lambda=1.2$ GeV, $g_{\sigma \Lambda_c \Lambda_c}$=7.3 and $g_{\omega \Lambda_c \Lambda_c}$=10.57. The result is performed by expending the basis function with N=120.}\label{fig2}
\end{figure}

The eigenvalues of the transformed Hamiltonian $H_\theta$ are drawn in Fig.\ref{fig2}. We can clearly see that all the eigenvalues of $H_\theta$ have three parts: the dark blue square, deep yellow triangle and green circle represent the bound state, resonant state and continuum, respectively. The bound state locates on the negative energy axis, while the continuous spectrum rotates clockwise with the angle $2\theta$, and the resonant state in the lower half of the complex energy plane, which is surrounded by the positive energy axis and the rotated continuum line and become isolated.

\begin{figure}
\centering
\includegraphics[width=3.13in, keepaspectratio]{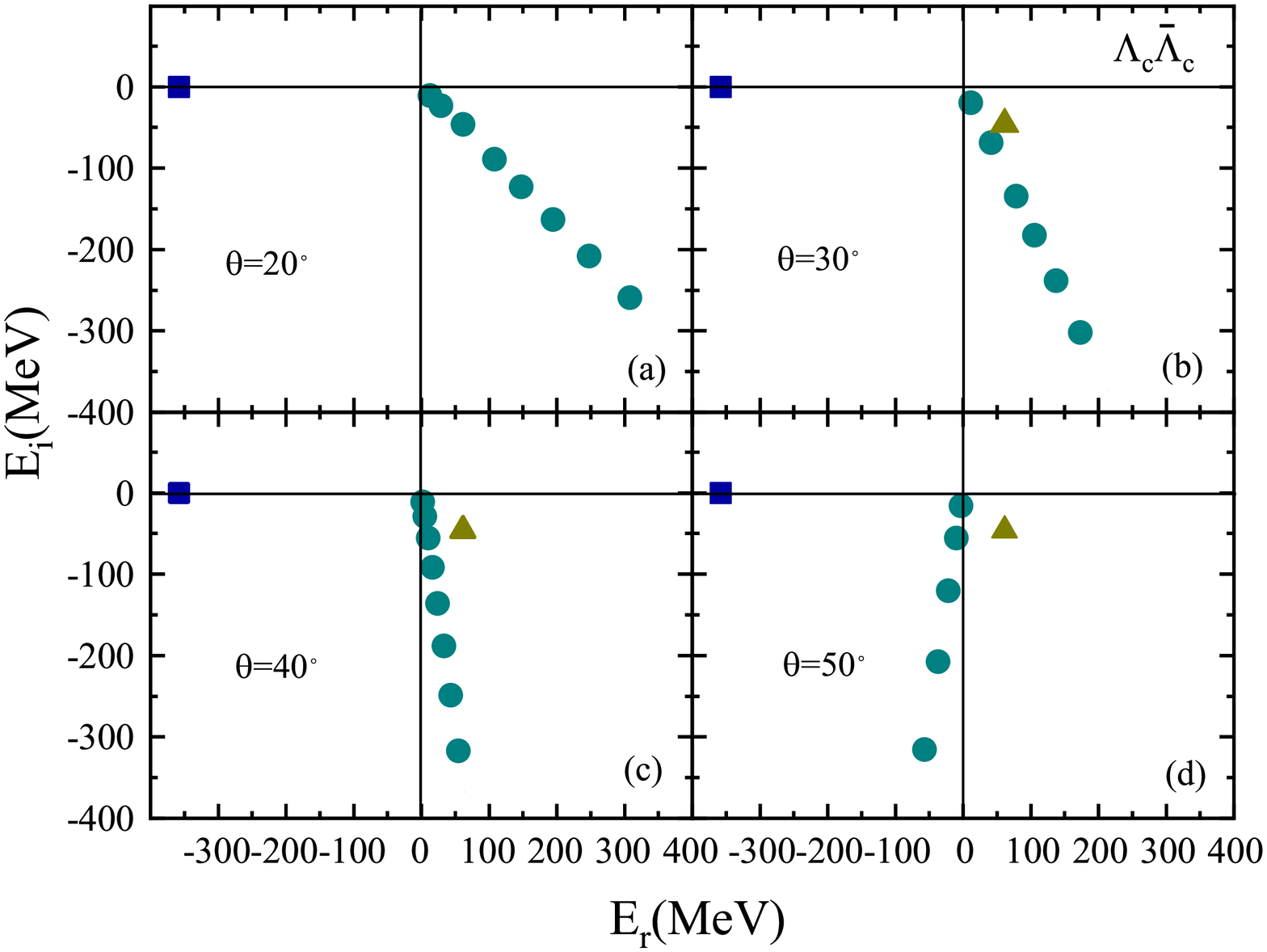}\\
\caption{(Color online) The resonant and continuous spectra varying with the complex rotation angle in the complex energy plane.
Except for the complex rotation angle, other parameters are set same as those in Fig. \ref{fig2}.}\label{fig3}
\end{figure}

In order to show how the resonant state is separated from the continuum by complex rotation, the eigenvalues of $H_\theta$ with different complex scale angle $\theta$ are plotted in Fig.\ref{fig3}, and other parameters are the same as those in Fig.\ref{fig2}. In Fig.\ref{fig3}(a),
when $\theta = 20^{\circ}$, we can only see the continuous spectrum, it is quite difficult to observe the resonant state in the complex energy plane.
In Fig.\ref{fig3}(b), when the rotation angel increases to $\theta = 30^{\circ}$, the resonant state begins to be separated from the continuum spectrum gradually. When $\theta = 40^{\circ}$, it is obviously that the resonant state is completely separated from the continuum in Fig.\ref{fig3}(c).
On the whole, from Fig.\ref{fig3}(b) to Fig.\ref{fig3}(d), no matter how the complex scale angle rotates, the position of the resonant state is almost unchanged in the complex energy plane. The above results suggest that the resonant state can be determined as long as the selected rotation angle is large enough.

\begin{figure*}
\centering
\includegraphics[width=6.5in, keepaspectratio]{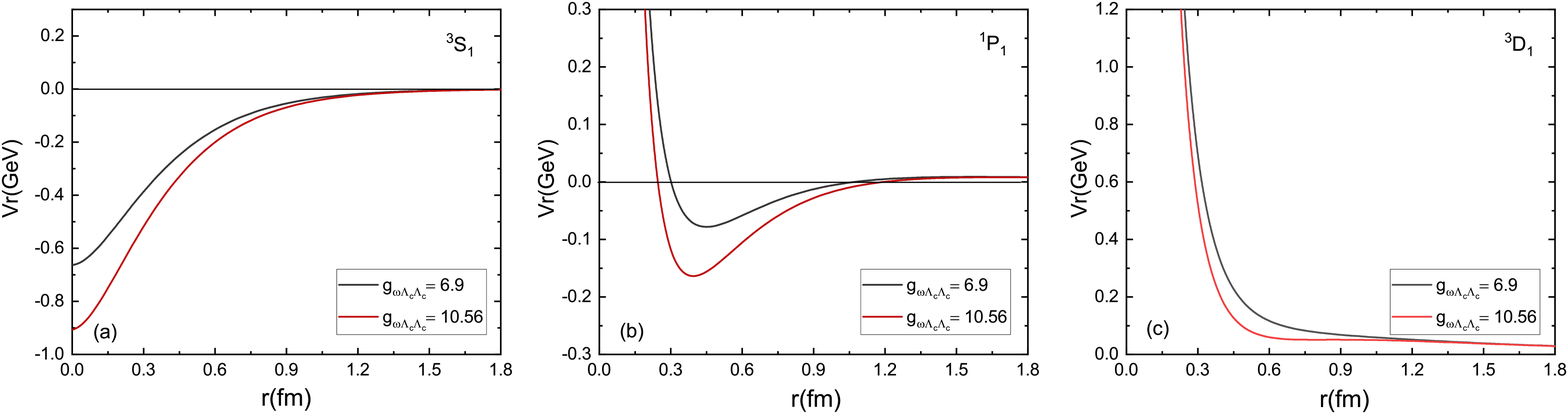}\\
\caption{(Color online) The total potentials in one boson exchange model with different orbit angular momentum L = 0, 1,2 for the $\Lambda_c\bar{\Lambda}_c$ system dependence on the r, respectively. The cutoff $\Lambda$ is set as 1.1 GeV, the value of coupling constant $g_{\omega}^{q}$ are set as 3.45 in the Nijmegen model, and 5.28 in the Bonn model \cite{Rijken:1998yy}, $g_{\sigma}^{q}$ is set as $g_{\sigma}^{q}$= 3.65, respectively. The relation between $g_{\omega/\sigma \Lambda_c \Lambda_c}$and $g_{\omega/\sigma}^{q}$ is $g_{\omega/\sigma \Lambda_c \Lambda_c} =2 g_{\omega/\sigma}^{q}$.}\label{fig4}
\end{figure*}

The total potentials of the $\Lambda_c\bar{\Lambda}_c$ system in one boson exchange model with the different orbit angular momentum L  =0, 1, 2 are ploted in Fig.\ref{fig4}. The black and red line correspond to the potential with $g_{\omega\Lambda_c \Lambda_c} =6.9$ and $g_{\omega \Lambda_c \Lambda_c} =10.56$ in Fig.\ref{fig4}, respectively. The centrifugal force term, $L(L+1)/2 \mu r^2$ is a repulsive force, the potential energy term provides attractive force, these two part are competitive. From Fig.\ref{fig4}(a), we can see that under the reasonable parameters, the potential becomes large enough to binding the two heavy baryons. In Fig.\ref{fig4}(b), The depth of the potential well becomes smaller, and a low potential barrier appears, compared with Fig.\ref{fig4}(a), which only has the center the potentials. In Table \ref{num2}, we find that when $g_{\omega\Lambda_c \Lambda_c} =6.9$, the energy and width of P resonant state is very small, and there is no P wave resonant for the $\Lambda_c\bar{\Lambda}_c$ system when $g_{\omega \Lambda_c \Lambda_c} =10.56$. In Fig.\ref{fig4}(c), the total potentials are larger than that in Fig.\ref{fig4}(b), they are relatively easy to form resonant states.

\renewcommand\tabcolsep{0.6cm}
\renewcommand{\arraystretch}{1.8}
\begin{table*}[!htbp]
  \caption{ The energy and width of bound and resonant states for the $\Lambda_c\bar{\Lambda}_c$ system. $E$ and $\Gamma$ represent the energy and width of resonant states in units of MeV, respectively. The cutoff $\Lambda$ is set as 1.1 GeV. The value of coupling constant $g_{\omega}^{q}$ are set as 3.45 in the Nijmegen model, and 5.28 in the Bonn model \cite{Rijken:1998yy}, $g_{\sigma}^{q}$ is set as $g_{\sigma}^{q}$= 3.65, respectively. The relation between $g_{\omega/\sigma \Lambda_c \Lambda_c}$ and $g_{\omega/\sigma}^{q}$ is $g_{\omega/\sigma \Lambda_c \Lambda_c} =2 g_{\omega/\sigma}^{q}$. The notation $\ldots$ stands for no bound or resonant state solutions.}\label{num2}
  \begin{tabular}{c|cccccc}\toprule[2pt]
     {\,$g_{\omega \Lambda_c \Lambda_c}$\,}

                        &$L$   &$E$      &$\Gamma$                         &$L$   &$E$      &$\Gamma$\\\hline
                        &$^1S_0$    &-131.66     &\ldots                   &$^3D_1$    &38.86       &183.82\\

                        &$^3S_1$   &-126.13      &\ldots                   &$^1D_2$   &32.71        &192.12\\

        6.9             &$^3P_0$   &7.15         &8.28                     &$^3D_2$   &34.94        &188.42\\

                        &$^1P_1$    &9.16        &13.82                    &$^3D_3$   &28.78        &193.7\\

                        &$^3P_2$    &9.90       &17.0      \\\hline

                        &$^1S_0$    &-240.84     &\ldots                   & $^3D_1$    &54.61       &131.8\\

                        &$^3S_1$    &-224.53     &\ldots                   & $^1D_2$   &49.66        &152.64\\

      10.56             &$^3P_0$    &\ldots      &\ldots                   & $^3D_2$   &51.23        &144.34\\

                        &$^1P_1$    &\ldots      &\ldots                   & $^3D_3$   &44.79        &160.1\\

                        &$^3P_2$    &\ldots      &\ldots       \\\hline

  \bottomrule[2pt]
  \end{tabular}
\end{table*}

We extend the complex scale method to solve the Schr\"{o}dinger equation and obtain the bound state and the resonant state for $\Lambda_c\bar{\Lambda}_c$ system numerically. The solution of resonant state has the form $E-i\Gamma /2$ , where $E$ is the resonance energy and $\Gamma$ is its decay width. The coupling strength $g_{\omega \Lambda_c \Lambda_c}$ is twice as much as $g_{\omega}^{q}$ in heavy quark effective theory. The coupling constant $g_{\omega}^{q}$ are set as 3.45 in the Nijmegen model, and 5.28 in the Bonn model as input benchmark parameters. We can obtain the energies of bound states, the energies and widths of the resonant states for the $\Lambda_c\bar{\Lambda}_c$ system with different angular momentum $L$, which are listed in Table \ref{num2}.

In Table \ref{num2}, we find that there exists S wave bound state in each case with $\Lambda=1.1$ GeV, whose binding energy is about hundreds MeV, and the difference of energies between the $^1S_0$ and $^3S_1$ state is about ten MeV. For the case $g_{\omega \Lambda_c \Lambda_c}=6.9$, the $\Lambda_c\bar{\Lambda}_c$ system can form P wave resonant states, the energies and widths are about several to more than a dozen MeV. In addition, the $\Lambda_c\bar{\Lambda}_c$ system can also form the D wave resonant states, the energies are dozens of MeV and widths are more than 100 MeV. The resonance widths increases with the increasing of angular momentum L for $\Lambda_c\bar{\Lambda}_c$ system, which indicates that the resonant states become more and more unstable. We find that the quantum number of the resonant state $^3D_1$ agrees with the quantum number $J^{PC}=1^{--}$ of Y(4630). Then, We want to know whether there are any suitable parameter consistent with the energy and width of Y(4630).

\begin{figure}
\centering
\includegraphics[width=3.10in, keepaspectratio]{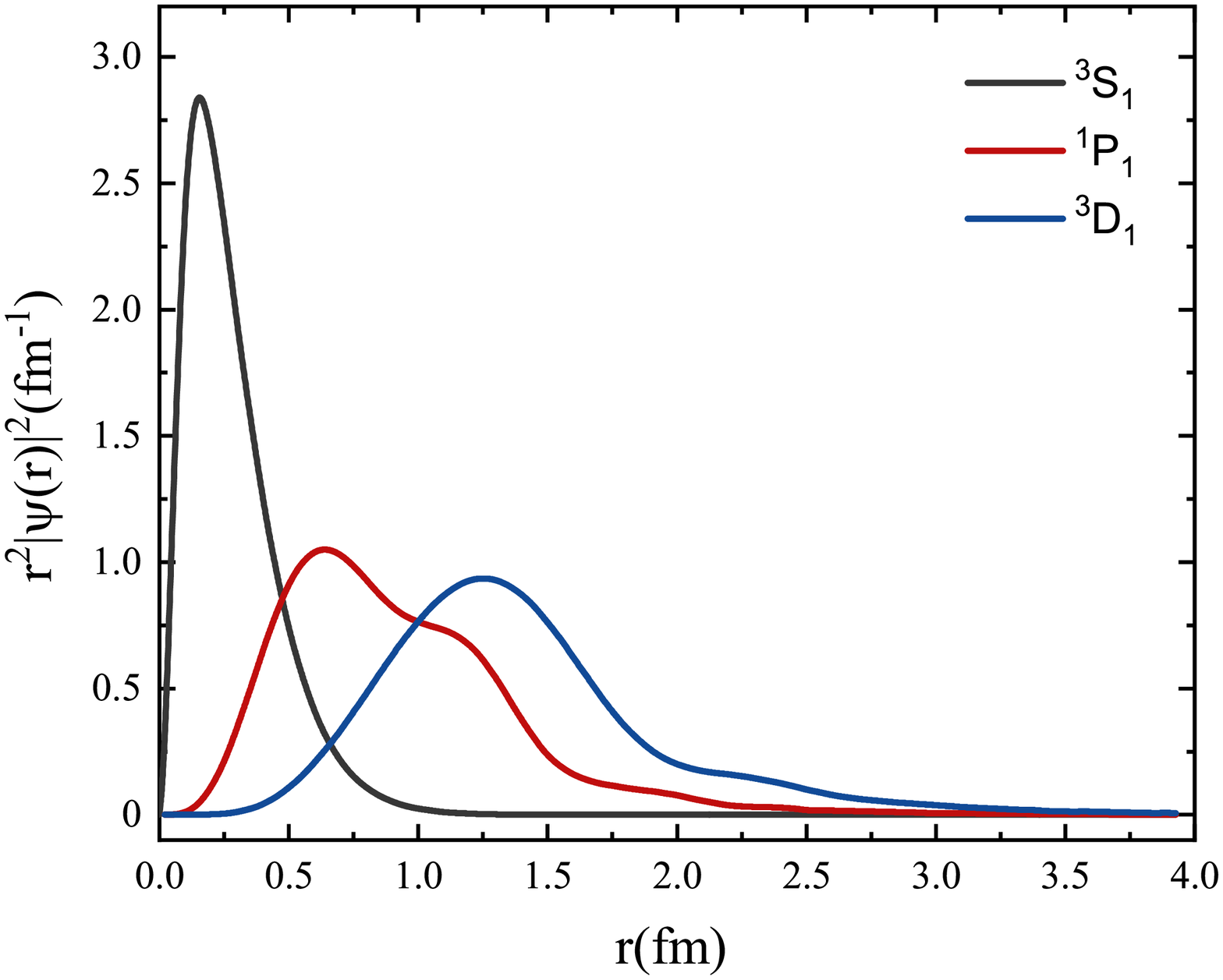}\\
\caption{(Color online) Radial density distributions in the coordinate space for the bound state $^3S_1$, and the resonant states $^1P_1$, $^3D_1$ with $g_{\sigma \Lambda_c \Lambda_c}=7.3$, $g_{\omega \Lambda_c \Lambda_c}=10.56$, and $\Lambda=1.1$ GeV. }\label{fig5}
\end{figure}

In Fig.\ref{fig5}, we show the radial density distributions for the bound state $^3S_1$, and the resonant states $^1P_1$, and $^3D_1$ with the coupling constants $g_{\sigma \Lambda_c \Lambda_c}=7.3$, $g_{\omega \Lambda_c \Lambda_c}=10.56$, the cutoff parameter $\Lambda=1.1$ GeV. The black, red and blue lines represent bound state $^3S_1$ ,and resonant states $^1P_1$, $^3D_1$, respectively. It can been seen that the black bound state converges when the radius around 0.25 fm, but the red and blue resonant states begin to converge when the radius about 0.75 and 1.25 fm, which indicate that compared with the bound state, the resonant states are more dispersed.

\begin{figure}
\centering
\includegraphics[width=3.10in, keepaspectratio]{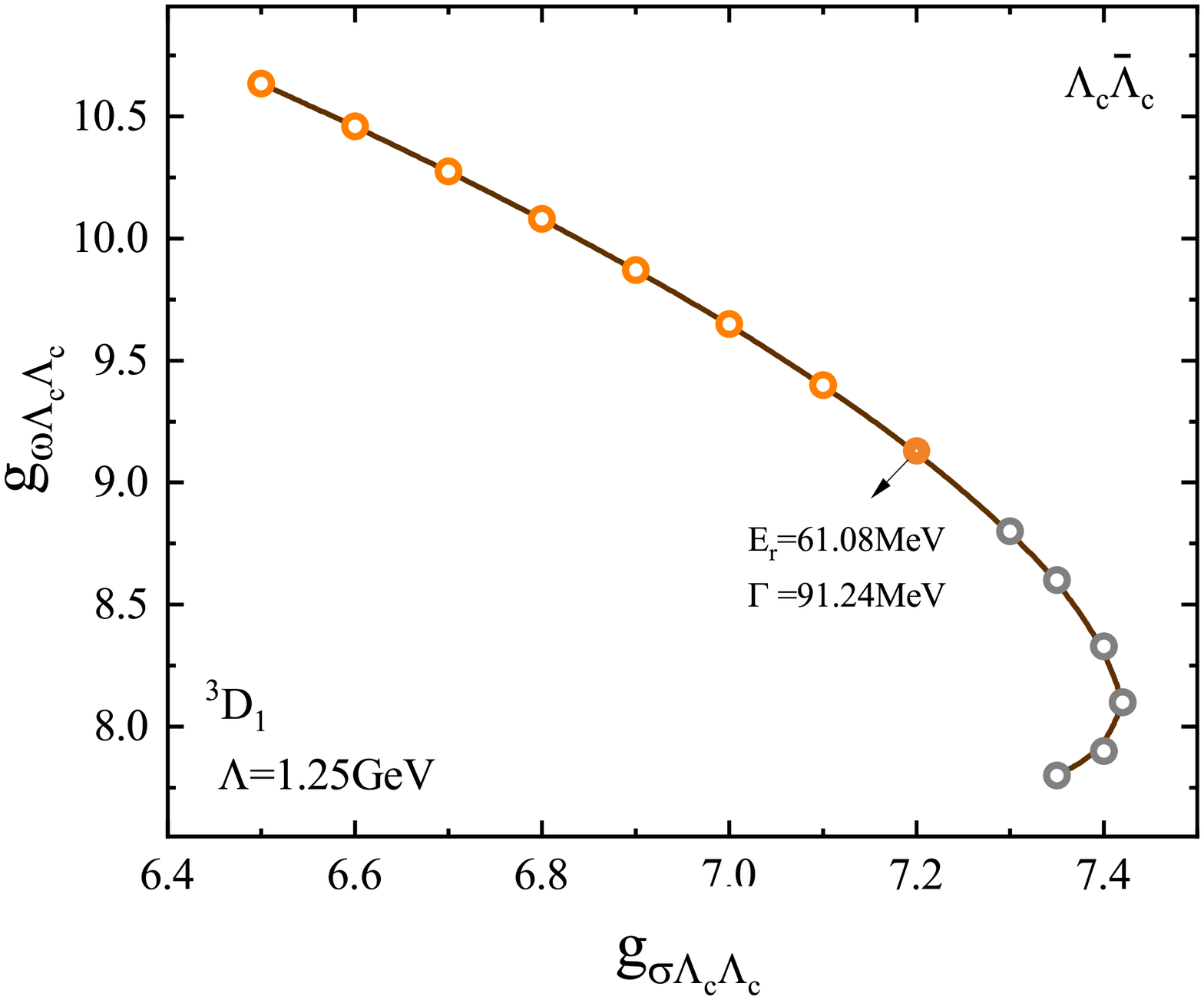}\\
\caption{(Color online) The values of $g_{\omega \Lambda_c \Lambda_c}$ and $g_{\sigma \Lambda_c \Lambda_c}$ for the decay width of $\Lambda_c\bar{\Lambda}_c$ system varying from 60 to 133 MeV, with $\Lambda=1.25$ GeV. }\label{fig6}
\end{figure}

The coupling constants $g_{\sigma \Lambda_c \Lambda_c}$ and $g_{\omega \Lambda_c \Lambda_c}$ in the Lagrangians are difficult to extract from experiments.
In Ref.\cite{Lee:2011rka}, the coupling constants of the heavy charmed baryons and light mesons can be approximately determined by nucleon-meson coupling the obtained the numerical values $g_{\sigma \Lambda_c \Lambda_c}$=5.64 and $g_{\omega \Lambda_c \Lambda_c}$=10.57. However, the coupling constant $g_{\sigma \Lambda_c \Lambda_c}$ is 7.3 in a $\sigma$ model \cite{Riska:1999fn}, the coupling constant $g_{\omega \Lambda_c \Lambda_c}=6.9$ in the Nijmegen model, and $g_{\omega \Lambda_c \Lambda_c}=10.56$ in the Bonn model \cite{Rijken:1998yy}. Therefore, the value of these two coupling strengths have large uncertainties. Taking into account the width uncertainties from 60 to 133 MeV of Y(4630), we show the different values of $g_{\omega \Lambda_c \Lambda_c}$ and $g_{\sigma \Lambda_c \Lambda_c}$ which satisfy the mass of Y(4630) in Fig.\ref{fig6}. When $g_{\sigma \Lambda_c \Lambda_c}= 7.2$, and $g_{\omega \Lambda_c \Lambda_c}=9.13$, there is a resonant state for $\Lambda_c\bar{\Lambda}_c$ system, with the energy 61.08 MeV, and width 91.24 MeV.
The width of X(4630) reported by Belle Collaboration is its total width, the $\Lambda_c\bar{\Lambda}_c$ is just one of its partial decay channels, thus
the decay width of the $\Lambda_c\bar{\Lambda}_c$ resonance state should be less than the width of X(4630). In Fig.\ref{fig6}, the orange dots represents the width less than the width 92 MeV of Y(4630), while the gray dots represent the width greater than 92 MeV. We can see that the range covered by the orange dots can satisfy the requirements of the experiment.

\begin{figure}
\centering
\includegraphics[width=2.8in, keepaspectratio]{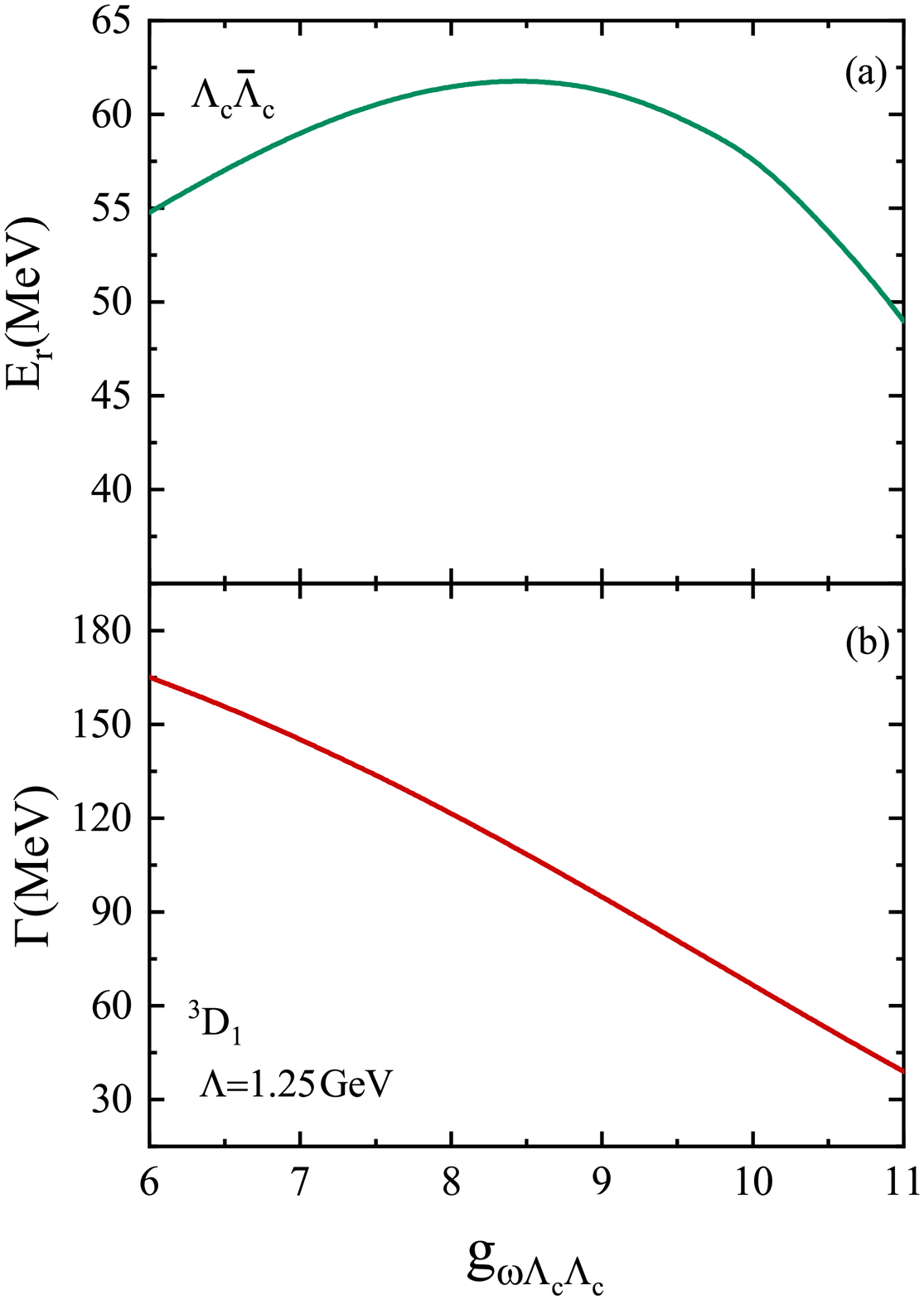}\\
\caption{(Color online) The variation of energy and width of $^3D_1$ wave resonant state with the change of coupling constant $g_{\omega \Lambda_c \Lambda_c}$ for $\Lambda_c\bar{\Lambda}_c$ system with $\Lambda=1.25$ GeV . }\label{fig7}
\end{figure}

In order to clear the dependence of energy and width on the coupling strength $g_{\omega \Lambda_c \Lambda_c}$, we present the energy and width as a function of coupling strength $g_{\omega \Lambda_c \Lambda_c}$ for resonant state $^3D_1$ in Fig.\ref{fig7}. As we can see, in Fig.\ref{fig7}(a), the energy of the resonant state increases slowly and then gradually decreases with the increasing of $g_{\omega \Lambda_c \Lambda_c}$, and reaches a maximum value when $g_{\omega \Lambda_c \Lambda_c}$ is about 9.0. Unlike the change of the energy, in Fig. \ref{fig7}(b), the width decreases significantly with the change of the coupling constant $g_{\omega \Lambda_c \Lambda_c}$.

\begin{figure}
\centering
\includegraphics[width=2.8in, keepaspectratio]{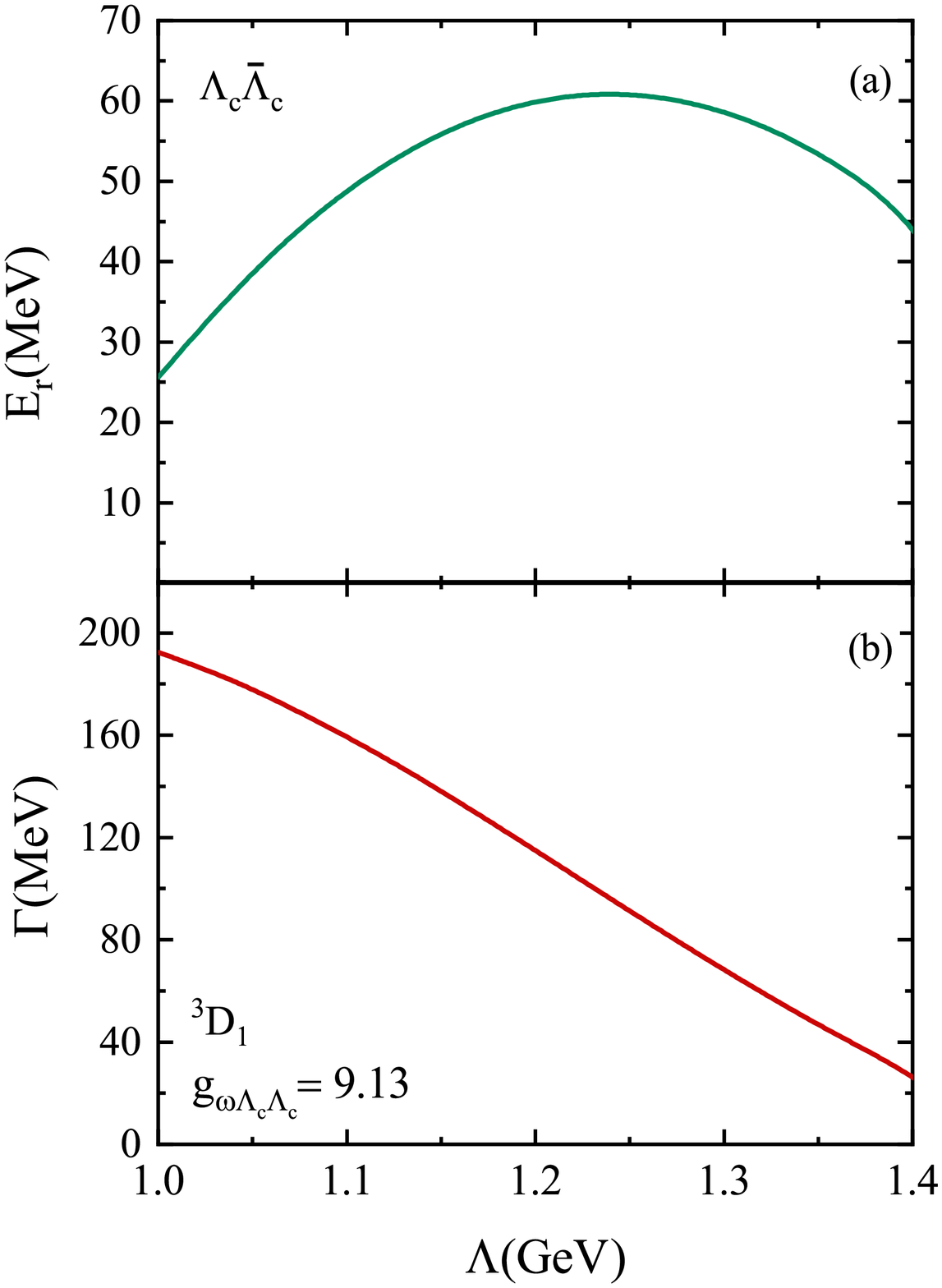}\\
\caption{(Color online) The energy and width of $^3D_1$ wave resonant state as a function of the cutoff parameter $\Lambda$ for $\Lambda_c\bar{\Lambda}_c$ system with $g_{\omega \Lambda_c \Lambda_c}=9.13$. }\label{fig8}
\end{figure}

\begin{figure}
\centering
\includegraphics[width=2.8in, keepaspectratio]{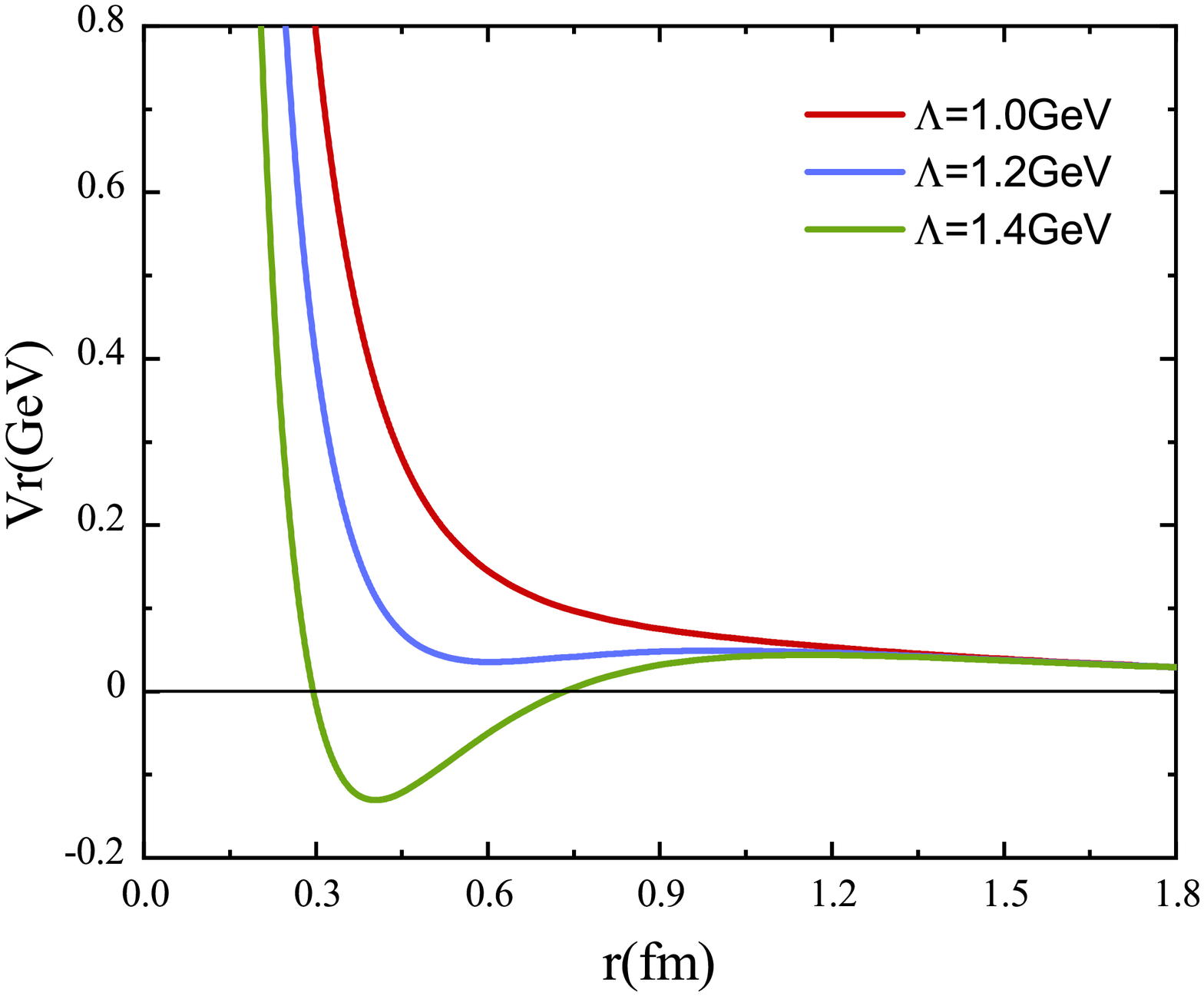}\\
\caption{(Color online) The total potentials in one boson exchange model with different cutoff parameter $\Lambda$ for $^3D_1$ wave resonant state dependence on the r. The coupling constants are set as $g_{\sigma \Lambda_c \Lambda_c}= 7.2$, and $g_{\omega \Lambda_c \Lambda_c}=9.13$. }\label{fig9}
\end{figure}

The cutoff parameter $\Lambda$ is related to the size of hadrons, and has a significant impact on the results of energy and width. For nucleon-nucleon interaction, the cutoff parameter $\Lambda$ is usually from 0.8 to 1.5 GeV, for the heavy hadron state, this value should be slightly larger. In Fig.\ref{fig8}(a) and (b), we present the energy and width of resonant state $^3D_1$ for $\Lambda_c\bar{\Lambda}_c$ system as a function of the $\Lambda$. From Fig.\ref{fig8}(a), we can see that the energy of resonant state $^3D_1$ increases from 25 MeV to 60 MeV, and then decreases to 40 MeV, when the cutoff parameter $\Lambda$ varies from 1.0 to 1.4. The corresponding width decreases from 200 MeV down to 20 MeV, as shown in Fig.\ref{fig8}(b). The energies and widths are related to the potential functions, which depend on the coupling constants and cutoff parameter. The dependence of potential function on coupling constants and cutoff is similar, thus we present the potential functions for the different cutoff parameters in Fig.\ref{fig9}. We can see that when the cutoff parameter becomes larger, the depth of potential well becomes deeper, the barrier effect is more obvious, the formed resonance state is more stable, and the decay width is smaller.

\renewcommand\tabcolsep{0.6cm}
\renewcommand{\arraystretch}{1.8}
\begin{table*}[!htbp]
  \caption{ The energy and width of bound and resonant states for $\Lambda_c\bar{\Lambda}_c$ systems, when the cutoff $\Lambda$ is 1.25 GeV, the coupling constant are set as $g_{\sigma \Lambda_c \Lambda_c}= 7.2$, and $g_{\omega \Lambda_c \Lambda_c}=9.13$. $E$ and $\Gamma$ represent the energy and width of resonant states in units of MeV, respectively.  The notation $\ldots$ stands for no bound or resonant state solutions.}\label{num3}
  \begin{tabular}{ccccccc}\toprule[2pt]

   &$L$   &$E$      &$\Gamma$                         &$L$   &$E$      &$\Gamma$\\\hline
   &$^1S_0$    &-408.9     &\ldots                   &$^3D_1$    &61.08       &91.24\\
   &$^3S_1$   &-376.27      &\ldots                   &$^1D_2$   &60.41        &125.06\\
   &$^3P_0$   &\ldots        &\ldots                    &$^3D_2$   &60.69       &112.68\\
   &$^1P_1$    &\ldots       &\ldots                    &$^3D_3$   &55.99       &140.98\\
   &$^3P_2$    &\ldots      &\ldots      \\\hline

  \bottomrule[2pt]
  \end{tabular}
\end{table*}

In Table \ref{num3}, we list the bound and resonant states for the $\Lambda_c\bar{\Lambda}_c$ systems, when the cutoff $\Lambda$ is 1.25 GeV, the coupling constants are set as $g_{\sigma \Lambda_c \Lambda_c}= 7.2$, and $g_{\omega \Lambda_c \Lambda_c}=9.13$. From the Table \ref{num3}, we can see that if $^3D_1$ resonant state is a reasonable explanation for Y(4630), there may also exist $^1D_2$, $^3D_2$ and $^3D_2$ resonant states around Y(4630), $^1S_0$ and $^3S_1$ deeper bound states, but cannot form P resonant state. These states can be investigated in future experiments. Although the angular momentum of the P state is lower than that of the D state, the height of potential barrier is smaller, and it is not easier to form a resonance state than the D state. It can be seen that the P state can be formed when $g_{\omega \Lambda_c \Lambda_c}$=6.9 in Table \ref{num2}, and there are no resonance state when $g_{\omega \Lambda_c \Lambda_c}$=10.56 in Table \ref{num2} and in Table \ref{num3}. Therefore, the P-wave resonance state can be formed within a certain parameter range. The P-wave resonance state is not observed in the experiment, probably because it not form a resonance state, or the researchers need further to analyze more data.

\section{Summary}\label{sec4}
Recent year, many new exotic hadrons have been discovered, some of them can be explained by hadronic molecular states in one-boson-exchange model.
In this paper, we have investigated the $\Lambda_c\bar{\Lambda}_c$ system using complex scaling method in one boson exchange model. The numerical results indicated that the $\Lambda_c\bar{\Lambda}_c$ system can form not only S wave bound state, but also higher angular momentum L resonant states. When the coupling constants are taken as $g_{\sigma \Lambda_c \Lambda_c}=7.2$ and $g_{\omega \Lambda_c \Lambda_c}=9.13$, there exists a $^3D_1$ wave resonant state with energy $E=61.08$ MeV and decay width $\Gamma=91.24$ MeV, which is consistent with the exotic hadron state Y(4630) with the quantum number $J^{PC}=1^{--}$, mass 4634 MeV and width 91.24 MeV. If $^3D_1$ resonant state is a reasonable explanation for Y(4630), there may also exist other bound and resonant states around Y(4630), which remains to be verified experimentally in future.

\vfill

\section*{ACKNOWLEDGMENTS}
This work was supported in part by the National Natural Science Foundation of China (No.11935001), the Natural Science Foundation of Anhui Province (No.2108085MA20, No.2208085MA10), and the Key Research Foundation of Education Ministry of Anhui Province of China (No.KJ2021A0061).

\end{document}